\documentclass[pra,twocolumn,showpacs,amsmath,amssymb]{revtex4}

\usepackage{graphicx}
\usepackage{epsfig}
\usepackage{amsmath}
\usepackage{graphics}
\usepackage{color}

\begin{document}

\title{Excitation Transport through a Domain Wall in a Bose--Einstein Condensate}

\author{Shohei Watabe$^{1,2}$} 
\altaffiliation{Present Address: Department of Physics, The University of Tokyo, Tokyo 113-0033, Japan}
\author{Yusuke Kato$^{3}$}
\author{Yoji Ohashi$^{1,2}$}
\affiliation{$^{1}$ 
Department of Physics, Keio University, 3-14-1 Hiyoshi, Kohoku-ku, Yokohama 223-8522, Japan}
\affiliation{$^{2}$ 
CREST(JST), 4-1-8 Honcho, Kawaguchi, Saitama 332-0012, Japan}
\affiliation{$^{3}$ 
Department of Basic Science, The University of Tokyo, Tokyo 153-8902, Japan}

\begin{abstract} 
We investigate the tunneling properties of collective excitations through a domain wall in the ferromagnetic phase of a spin-1 spinor Bose--Einstein condensate. Within the mean-field theory at $T=0$, we show that the transverse spin wave undergoes perfect reflection in the low-energy limit. This reflection property differs considerably from that of a domain wall in a Heisenberg ferromagnet where spin-wave excitations exhibit perfect transmission at arbitrary energy.
When the Bogoliubov mode is scattered from this domain wall soliton, the transmission and reflection coefficients exhibit pronounced non-monotonicity.
In particular, we find perfect reflection of the Bogoliubov mode at energies where bound states appear.
This is in stark contrast to the perfect transmission of the Bogoliubov mode with arbitrary energy through a dark soliton in a scalar Bose--Einstein condensate.
\end{abstract}

\maketitle

\section{Introduction}

Bloch and N\'eel walls in ferromagnetic materials are domain walls with well-known structures in condensed matter, where spin configurations are twisted in a magnetic domain region with an angular displacement of $180^\circ$~\cite{Kittel1996}. 
It is desirable to use domain wall motions in ferromagnets in the aim of their application to memory devices~\cite{Slonczewski1996,Grollier2002,Klaui2003,Grollieri2003,Vernier2004,Tatara2004,Yamaguchi2004,Zhang2004,Thiaville2005,Klaui2005,TataraReview,Ieda2010}. 
For example, current-driven domain wall motion associated with spin/momentum transfer from electrons to the domain wall has been theoretically discussed~\cite{Tatara2004} and has been experimentally observed~\cite{Yamaguchi2004}. 

\begin{figure}[tbp]
\includegraphics[height=2.8cm,width=6cm]{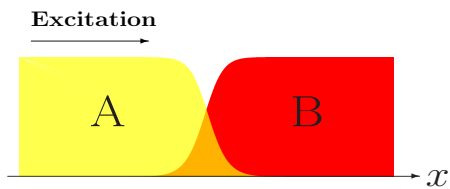}
\caption{(color online) One-dimensional tunneling of excitations through a domain wall of a Bose--Einstein condensate. For a ferromagnetic spin-1 Bose--Einstein condensate, A and B represent different hyperfine states of the condensate (A corresponds to the $S_{z}=-1$ state and B to $S_{z}=+1$). A right-moving incident excitation is scattered from the domain wall.
}
\label{fig1} 
\end{figure} 

Domain walls are also found in two-component Bose--Einstein condensates and spin-1 Bose--Einstein condensates. For spin-1 Bose--Einstein condensates, spin domain structures~\cite{Stenger1998,Miesner1999}, tunneling from metastable spin domains~\cite{Kurn1999}, and the formation of spin textures and domains~\cite{Sadler2006,Vengalattore2010} have been experimentally investigated.
An earlier theoretical paper~\cite{Isoshima1999} reported that ferromagnetic domain walls in a spin-1 Bose--Einstein condensate spread out over the entire spatial region of the system.
However, a domain wall may exhibit solitary behavior with a size comparable with the healing length of the Bose--Einstein condensate, 
which is much smaller than the system size. 
In addition, a uniform particle density around the domain wall was reported in Ref.~\cite{Isoshima1999}, 
which is analogous to that of the Heisenberg model, where the magnitude of the spin vector is spatially uniform. However, gaseous ferromagnetic spin-1 Bose--Einstein condensates may be free from such a constraint, 
so that their particle/spin density may become non-uniform in the domain wall region. 
A search for such a domain is anticipated to reveal
a solitary domain wall in spin-1 Bose--Einstein condensates.

Furthermore,
transport phenomena in gaseous ferromagnetic Bose--Einstein condensates are much less understood than those in solid materials. 
Scattering between low-lying modes and a domain wall in a Bose--Einstein condensate will be interesting, 
if it exhibits different characters from conventional ferromagnets. 
In contrast to the Heisenberg model case, 
the spin density vector amplitude may be spatially non-uniform in spinor Bose--Einstein condensates. 
This character is expected to provide interesting insights into ferromagnets and spinor Bose--Einstein condensates. 
In addition, since these scattering problems can be expressed in terms of Nambu--Goldstone modes and topological excitations, 
this concept is not specific to cold atomic gases, but can also be applicable to other symmetry-broken states.

We investigate the tunneling properties of collective excitations through a ferromagnetic domain wall in a spin-1 Bose--Einstein condensate. 
We start by reconsidering  the domain structure of the ferromagnetic Bose--Einstein condensate with a given boundary condition.
For simplicity, we consider a three-dimensional system with a planar domain wall which spatially varies in the $x$-direction. 
We study one-dimensional tunneling at $T =0$, as shown in Fig.~\ref{fig1}, 
where A and B correspond to different hyperfine states $S_{z} (= \pm 1)$ of the spin-1 Bose--Einstein condensate.

The main results of this paper are summarized as follows.
We find that the domain wall structure of the spin-1 Bose--Einstein condensate differs considerably from that in the ferromagnetic Heisenberg model.
Its transverse spin wave also exhibits strikingly different tunneling properties. 
The transverse spin wave in ferromagnets transmits perfectly through a domain wall~\cite{Yan2011}. 
In contrast, the transverse spin wave of the present Bose system is perfectly reflected by the ferromagnetic domain wall in the low-energy limit. For this perfect reflection,  the wave function does not vanish in the low-energy limit. 
This contrasts with a conventional single particle in quantum mechanics, 
which undergoes perfect reflection in the low-energy limit usually due to the absence of the wave function.
The quadrupolar spin wave in a spin-1 Bose--Einstein condensate exhibits conventional reflection properties in the long-wavelength limit
for which perfect reflection occurs due to the absence of the wave function in the long-wavelength limit. 
The coexistence of both perfect reflection and a non-vanishing wave function in the low-energy limit 
is due to the existence of a damping mode whose damping length becomes infinite in the low-energy limit.

Spin-1 Bose--Einstein condensates have another Nambu--Goldstone mode: the Bogoliubov excitation. 
In the low-energy limit, the Bogoliubov mode does not undergo perfect reflection when it is incident on the domain wall.
This is similar to the situation for a scalar Bose--Einstein condensate with a dark soliton. 
However, energy-dependent non-monotonic tunneling properties are observed at a finite energy. 
In particular, the perfect reflection occurs when the bound state appears at the domain. 
This differs from the dark soliton case in the scalar bosons, where the transmission coefficient is independent of energy.

This paper is organized as follows. In Sec.~\ref{SecII}, we compare the domain wall in a ferromagnetic spin-1 Bose--Einstein condensate with that in the ferromagnetic Heisenberg model. In Sec.~\ref{SecIII}, 
we examine the tunneling properties of excitations through the ferromagnetic domain wall. 

\section{Comparison with Domain Wall of Heisenberg Model}\label{SecII}

At $T=0$ in a three-dimensional system, the domain structure of a spin-1 Bose--Einstein condensate is described by the Gross--Pitaevskii equation~\cite{Ohmi1998,Ho1998}
for the condensate wave function ${\hat \Phi} = \left ( \Phi_{+1} , \Phi_0 , \Phi_{-1} \right )^{\rm T}$ (where the subscripts $\pm1, 0$ represent hyperfine states in the $S = 1$ spin state), 
\begin{eqnarray}
i\hbar
{\partial {\hat \Phi}({\bf r},t) \over \partial t}=
\left(
\begin{array}{ccc}
h_{+}({\bf r},t) & \displaystyle{c_1 \over \sqrt{2}} F_-& 0\\
\displaystyle{c_1 \over \sqrt{2}}F_+ & h({\bf r},t) & \displaystyle{c_1 \over \sqrt{2}}F_-\\
0 & \displaystyle{c_1 \over \sqrt{2}}F_+ & h_{-}({\bf r},t) \\
\end{array}
\right)
{\hat \Phi}({\bf r},t). 
\label{eq.3}
\end{eqnarray}
Here, we have introduced $h({\bf r},t) \equiv - \hbar^{2}\nabla^2/(2m) + c_{0} \rho({\bf r},t)$, 
$h_{\pm} ({\bf r},t) \equiv h({\bf r},t) \pm c_1F_z ({\bf r},t)$ and $F_\pm \equiv F_x\pm iF_y$, 
where $m$ is the atomic mass, $\rho({\bf r},t)={\hat \Phi}^\dagger({\bf r},t){\hat \Phi}({\bf r},t)$ is the particle density,
and ${\bf F}={\hat \Phi}^\dagger({\bf r},t){\bf S}{\hat \Phi}({\bf r},t)$ is the spin density. 
The spin quantization axis of $S=1$ spin matrices ${\bf S}=(S_x,S_y,S_z)$ is chosen to be parallel to the $z$-axis. 
The two coupling constants $c_0 = 4\pi\hbar^{2}(a_{0}+2a_{2})/(3m)$ and $c_1 = 4\pi\hbar^{2}(a_{2}-a_{0})/(3m)$ are for spin-independent and spin-dependent interactions, respectively~\cite{Ho1998}, 
where $a_{S}$ is the s-wave scattering length for the total spin $S = 0$ or $2$ channel. 

\begin{figure}[tbp]
\includegraphics[width=8cm]{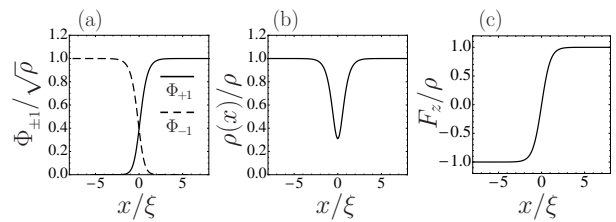}
\caption{
Calculated (a) condensate wave functions, (b) particle density, and (c) spin density of $F_{z}$ 
as a function of $x$. $\Phi_{0}$ is absent everywhere. 
$\xi$ is the healing length and is given by $\xi = \hbar/\sqrt{mc_{+}\rho}$.
}
\label{fig2} 
\end{figure} 

We consider a ferromagnetic spin-dependent interaction  ($c_{1} < 0$) 
and a planar ferromagnetic domain wall with only an $x$ dependence (i.e., $F_{z} (x=\pm \infty) = \pm \rho$, where $\rho$ is the density at $|x| = \infty$).
The condensate wave function for the stationary state is determined from (\ref{eq.3}) by setting ${\hat \Phi}(x,t)=e^{-i\mu t/\hbar}{\hat \Phi}(x)$, where $\mu$ is the chemical potential, under the boundary condition ${\hat \Phi} (x=-\infty) = (0,0,\sqrt{\rho})^{\rm T}$ and ${\hat \Phi}(x=+\infty) = (\sqrt{\rho},0,0)^{\rm T}$. Figures~\ref{fig2}(a)--(c) respectively show the spatial profile of the condensate wave function, the particle density, and the spin density of $F_{z}$, which were obtained by solving the Gross--Pitaevskii equation with the chemical potential $\mu = (c_{0} + c_{1})\rho$. 
The result of Figs.~\ref{fig2}(a)--(c) is characterized by $\Phi_{0} = 0$. 

The domain wall shown in Fig.~\ref{fig2}(a) is quite different from that in a ferromagnet described by the Heisenberg model. 
In the continuum limit, this spin model is described by 
\begin{align}
H_{\rm spin} = \int \frac{d^{3}x}{a^{3}} 
\left [ \frac{J}{2} (\nabla \boldsymbol{\sigma})^{2} - \frac{K}{2} \sigma_{z}^{2} \right ], 
\label{Hspin}
\end{align}
where the easy axis is taken to be parallel to the $z$-axis. $\boldsymbol{\sigma} = (\sigma_{x}, \sigma_{y}, \sigma_{z})^{\rm T}$ is the spin vector, $a$ is the lattice constant, and $J (> 0)$ and $K(>0)$ are respectively the exchange coupling and easy axis anisotropy constants. 
A domain wall solution to (\ref{Hspin}) is given by 
$(\sigma_{x}, \sigma_{y}, \sigma_{z}) = (0,\sigma/\cosh(x/\lambda),\sigma\tanh(x/\lambda))$, where $\lambda \equiv \sqrt{J/K}$~\cite{TataraReview,Landau1960}. The magnitude of $\boldsymbol{\sigma}$ is homogeneous and the spin direction continuously changes around the magnetic wall. 
The present ferromagnetic domain wall for the spin-1 Bose--Einstein condensate differs markedly from this domain wall in the following two ways: 
(i) $F_{x,y} ( \propto |\Phi_{0}| )$ is absent because $\Phi_{0} = 0$ everywhere, even near the domain wall;
(ii) the magnitude of the spin density is not homogeneous and it vanishes at the center of the wall (see Fig.~\ref{fig2}(c)).

A domain wall structure with $F_{x,y} \neq 0$ has been also discussed in the system we are considering~\cite{Isoshima1999}. 
We find that energy of a domain wall in Ref.~\cite{Isoshima1999} with a non-zero value of $\Phi_0$ is higher than that shown in Fig.~\ref{fig2}.

The different domain wall structures between the Heisenberg model and this spin-1 Bose--Einstein condensate originate from the following causes.
For the spin-1 Bose--Einstein condensate, the spin-density interacts with itself (i.e., ${\bf F}^{2}(x)$) and the spatial structure is governed by the kinetic term of the condensate wave function. 
On the other hand, for the spatial structure of the domain wall in the Heisenberg model, the nearest-neighbor spin exchange interaction (i.e., $(\nabla \boldsymbol{\sigma})^{2}$ for the continuous approximation) in (\ref{Hspin}) is important. 

From the results in Fig.~\ref{fig2}, 
the condensate wave function can be simply expressed by 
$\hat{\Phi}(x) = ( \Phi_{+1}(x), 0, \Phi_{-1}(x) )^{\rm T}$, 
so that the Gross--Pitaevskii equation can be simplified as 
\begin{align}
\left [ - \frac{\hbar^{2}}{2m}\partial_x^2 -\mu + c_{+} |\Phi_{\pm 1}|^{2} + c_{-} |\Phi_{\mp 1}|^{2} \right ]\Phi_{\pm 1} = 0, 
\label{GPpm1}
\end{align}
where $c_{\pm} = c_{0} \pm c_{1}$. 
In this case, (\ref{GPpm1}) can be regarded as the two-component Gross--Pitaevskii equation,
if we regard $c_{+}$ and $c_{-}$ as the interaction parameters for the same and different species, respectively. 
In the ferromagnetic case, we have $c_{-} > c_{+}$, which corresponds to the condition for phase separation of the two-component Bose--Einstein condensate~\cite{Coen2001}. 
Figure~\ref{fig2}(a) clearly  shows the phase-separated domains of the two hyperfine states.

Even if we add the quadratic Zeeman term to the Gross--Pitaevskii equation 
to introduce the easy axis as the $z$-axis by reference to (\ref{Hspin}), 
the situation remains the same. 
For this quadratic Zeeman effect, the absence of $\Phi_{0}$ is energetically preferable~\cite{Kudo2010,QZ}. 
That is, the ferromagnetic spin-1 Bose--Einstein condensate has a domain wall solution even when no easy axis is introduced. 
In contrast, a ferromagnet in the Heisenberg model has a domain wall with finite thickness 
only when it has an easy axis (i.e., $K\neq 0$)~\cite{Kittel1996}.

\begin{figure}[tbp]
\includegraphics[width=3.5cm]{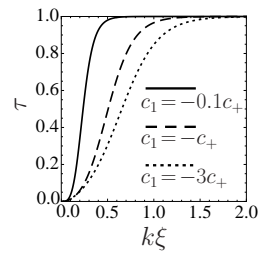}
\caption{
Calculated transmission probability $\tau$ through the domain wall 
as a function of the incident momentum when the transverse spin mode is incident on the domain wall.
}
\label{fig3} 
\end{figure} 

\section{Tunneling of Excitations through Domain Wall}\label{SecIII}

The tunneling properties of excitations can be determined by considering 
${\hat \Phi}(x,t)=\exp{(-i\mu t/\hbar)} ( {\hat \Phi(x)}+{\hat \phi}(x,t))$ 
(where ${\hat \phi} =(\phi_{+1},\phi_0,\phi_{-1})^{\rm T}$ describes fluctuations of the condensate wave function about the mean-field value) and retaining terms to $O({\hat \phi}(x))$ in the Gross--Pitaevskii equation (\ref{eq.3}). 
The ferromagnetic spin-1 Bose--Einstein condensate with $S_z = 1$ is known to have three kinds of collective modes~\cite{Ohmi1998,Ho1998}. 
Among them, the Bogoliubov mode is associated with phase fluctuations of the order parameter, 
where spin degrees of freedom is not crucial. 
The remaining two spin-wave excitations are associated with spin fluctuations in the $S_z = 0$ and $S_z = -1$ channels. 
While the former is called the transverse spin wave, characterized by $\delta F_\pm = \delta F_x \pm i \delta F_y = \sqrt{2} (\Phi_{\pm 1}^* \phi_0 + \Phi_{\mp 1} \phi_0^* )$, the latter is called the quadrupolar spin mode, 
characterized by $\delta Q_{+} = 2 \Phi_{+1}^* \phi_{-1}$ and $\delta Q_{-} = 2 \Phi_{+1} \phi_{-1}^*)$. 
We briefly note that the quadrupolar spin density is given by $Q_{\pm} \equiv \hat \Phi^\dag S_{\pm}^2 \hat \Phi = 2 \Phi_{\pm 1}^* \Phi_{\mp 1}$. 

After applying the Bogoliubov transformation, $\phi_{\pm 1,0} = u_{\pm 1,0} e^{-iEt/\hbar} - v_{\pm 1,0}^{*} e^{+iEt/\hbar}$, 
we can determine the reflection and transmission coefficients ($r$ and $\tau$, respectively) by solving the Bogoliubov equation under the boundary condition for the tunneling problem (see Appendix~\ref{Appendix}). 
Here, $r$ ($\tau$) is defined as the magnitude of  the ratio of the flux density of the reflected (transmitted) wave to that of the incident wave. 
Here, the flux density $J$ is given by $J =\sum_{j=\pm1, 0} {\rm Im}[u_{j}^{*}\partial_{x} u_{j} + v_{j}^{*}\partial_{x} v_{j}]/m$ and is 
proportional to the time-averaged energy flux~\cite{WatabeKato2011}. 

An excitation with $S_{z}=0$, corresponding to the transverse spin wave, 
is decoupled from other spin states (i.e., $\phi_{\pm 1}$), so that the transmission and reflection coefficients are obtained by solving the Bogoliubov equation for $S_{z} = 0$. 
Figure~\ref{fig3} shows the transmission coefficient $\tau$ of the transverse spin wave as a function of momentum. The excitation cannot pass through the domain wall in the low-momentum limit 
and the transmission coefficient increases monotonically with increasing momentum. 
(Perfect reflection in the low-momentum limit is proved in Appendix~\ref{Appendix2}.)

This perfect reflection contrasts with the case of the Heisenberg model (\ref{Hspin}).
In the latter case, the eigenfunction describing the spin wave in the presence of the domain wall is proportional to $[-ik\lambda + \tanh(x/\lambda)]\exp{(ikx)}$~\cite{TataraReview,Winter1961}. That is, the incident wave is perfectly transmitted through the domain wall for all $k$. 
This phenomenon in a magnetic nanowire has recently been studied by solving the Landau--Lifshitz--Gilbert equation~\cite{Yan2011}.  

The absence and presence of the amplitude of the order parameter at the domain wall are suggestive of the different tunneling properties of the transverse spin wave for the Heisenberg model and spin-1 Bose--Einstein condensates. 
However, a zero value of the order parameter at the domain wall is not related to perfect reflection. 
Let us consider the system where $\lim\limits_{x\rightarrow \pm \infty} (\Phi_{+1}, \Phi_{-1}) = (\sqrt{\rho},0)$ holds, 
but which has a region where $(\Phi_{+1}, \Phi_{-1}) = (0,\sqrt{\rho})$ holds around $x = 0$. 
In that system, we can find points where $\Phi_{+1} = \Phi_{-1}$, which leads to $|{\bf F}| = 0$. 
The absence of the spin density represents the same situation as that shown in Fig.~\ref{fig2}. 
In this case, however, the transverse spin wave exhibits perfect transmission in the low-energy limit (as proved in Appendix~\ref{Appendix2}). 
In these two cases, the excitation wave functions in the low-energy limit are given by the same form of the condensate wave functions.
In Bose--Einstein condensates, this property usually gives rise to perfect transmission 
of an excitation through a potential barrier in the low-energy limit~\cite{Kagan2003,Kato2008}.
For the present transverse spin wave, the damping length of the excitation solution becomes infinite in the low-energy limit. 
This makes possible for the damping mode of the excitation wave function to smoothly connect with the zero-energy mode corresponding to the condensate wave function in the low-energy limit. 
Consequently, depending on the configuration of the condensate wave function, 
the damping mode can remain alone in the transmission region where the condensate wave function is finite. 
See Appendix~\ref{Appendix2} for more details.

\begin{figure}[tbp]
\includegraphics[width=7cm]{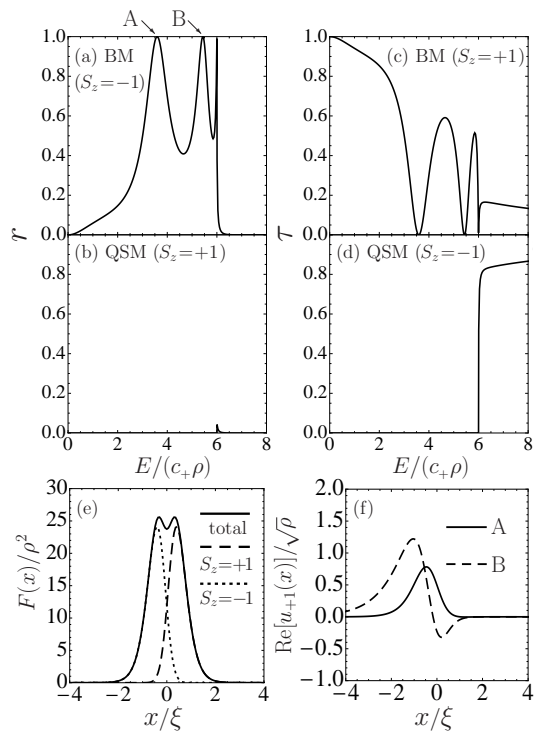}
\caption{
Transmission and reflection properties of excitations through the domain wall when the Bogoliubov mode is incident on the domain wall. 
Reflection probabilities of (a) the Bogoliubov mode (BM) ($S_{z} = - 1$) 
and (b) the quadrupolar spin mode (QSM) ($S_{z} = + 1$). 
Transmission probabilities of (c) the BM ($S_{z} = + 1$) 
and (d) the QSM ($S_{z} = - 1$). 
(e) A matrix element of the density spectral function in the low-energy regime ($E = 10^{-2}c_{+}\rho$). 
(f) The real part of the wave function $u_{+1}$, where the Bogoliubov mode is perfectly reflected.
The coupling constant is chosen to be $c_{1} = -3c_{+}$. 
}
\label{fig4} 
\end{figure} 

We now determine how the Bogoliubov mode and the quadrupolar spin mode are scattered from this domain wall. 
Hyperfine states $S_{z} = \pm 1$ are coupled as in (\ref{GPpm1}); 
however, for $x = -\infty$, the Bose--Einstein condensate of $S_{z} = -1$ only exists, so that 
an excitation of $S_{z} = -1$ corresponds to the gapless Bogoliubov mode with 
$E = [ \varepsilon (\varepsilon + 2c_{+}\rho )  ]^{1/2}$ 
and 
an excitation of $S_{z} = +1$ to the quadrupolar spin mode 
with an excitation gap that has the spectrum $E=\varepsilon + 2|c_{1}|\rho$, where $\varepsilon = \hbar^{2}k^{2}/(2m)$~\cite{Ho1998}.
For $x = \infty$, the Bose--Einstein condensate of $S_{z} = +1$ only exists, so that 
the roles for the two hyperfine states in the excitations are interchanged.

Figure~\ref{fig4} shows the reflection and transmission coefficients, $r$ and $\tau$, when the Bogoliubov excitation is scattered from the domain wall.
Intriguingly, reflection does not occur in the low-energy limit. 
Excitations in Bose--Einstein condensates are known to be perfectly transmitted through potential barriers in the low-energy limit; 
this phenomenon is referred to as anomalous tunneling~\cite{Kagan2003}. 
The Bogoliubov mode in the low-energy limit is a phase excitation of a Bose--Einstein condensate (i.e., a Nambu--Goldstone mode), 
which plays an essential role in anomalous tunneling~\cite{Takahashi2009,WKO2011}. 
However, the absence of reflection in the present case differs from anomalous tunneling. 
Figure~\ref{fig4}(e) shows the matrix element of the local density spectral function 
$F (x) \equiv |\Phi^{*} u - \Phi v|^{2}$~\cite{KatoWatabe2010} in the low-energy limit. 
In contrast to anomalous tunneling, 
the density modes of $S_{z} = \pm 1$ states localized at the domain wall appear even in the low-energy limit. 
(See Appendix~\ref{Appendix3} for more details, where these localized modes are discussed.)

With respect to scattering between a topological excitation and the Nambu--Goldstone mode, 
a dark soliton in a scalar Bose--Einstein condensate is known to be completely transparent to the Bogoliubov mode for all energies~\cite{Kovrizhin2001}. 
In this case, the localized density mode can also be found in the low-energy limit, 
given by $F(x) \propto \tanh ^{2}(x)/\cosh ^{4}(x)$ in the dimensionless form~\cite{Kovrizhin2001}, which is similar to the present case. 
However, in stark contrast to this dark soliton case, 
the transmission and reflection coefficients in the present case exhibit pronounced non-monotonicity. 
In particular, the Bogoliubov mode is perfectly reflected at some energy points (A and B in Fig.~\ref{fig4}(a)). 
This perfect reflection is strongly related to the bound state of the $S_{z} = +1$ state at the domain wall (see Fig.~\ref{fig4}(f)). 
This appears to be similar to resonance scattering, where the bound state increases the scattering cross-section~\cite{SakuraiBook}. 
However, the present bound state is specific to the ferromagnetic spin-1 Bose--Einstein condensate. 
For $x<0$, $(u_{+1},v_{+1})$ decays when the incident energy is lower than the energy gap of the quadrupolar spin mode. 
For $x > 0$, the wave functions $(u_{+1},v_{+1})$ behave as the Bogoliubov mode
whose damping mode is given by $\exp (- \kappa_{\rm B}x)$ for a positive energy $E(>0)$ (see also Appendix~\ref{Appendix} and \cite{kQ}). 
The damping structures on both sides of the domain wall are different.

For $E < 2|c_{1}|\rho$, the plane wave solution of the quadrupolar spin mode does not exist 
and the reflection and transmission of the quadrupolar spin mode are both forbidden (Figs.~\ref{fig4} (b) and (d)). 
When the incident Bogoliubov energy exceeds the energy gap of the quadrupolar spin mode (i.e., $E > 2|c_{1}|\rho$), 
the transmission coefficient of the quadrupolar spin mode ($S_{z} = -1$) suddenly increases (see Fig.~\ref{fig4}(d)).

\begin{figure}[tbp]
\includegraphics[width=7.3cm]{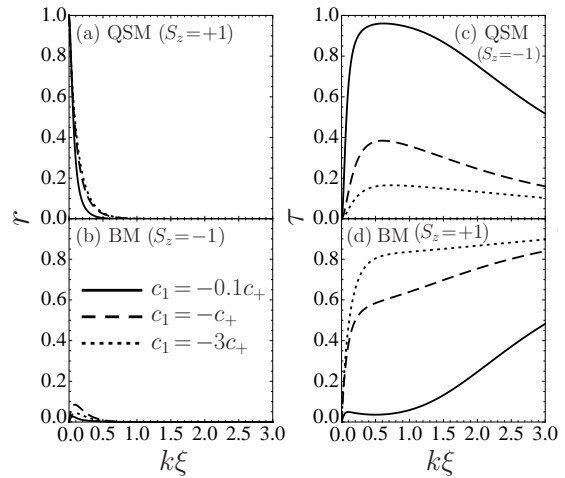}
\caption{
Transmission and reflection properties of excitations at a domain wall when the quadrupolar spin mode is incident on the domain wall. 
Reflection probabilities of (a) quadrupolar spin mode (QSM) ($S_{z} = + 1$) 
and (b) Bogoliubov mode (BM) ($S_{z} = - 1$). 
Transmission probabilities of (c) QSM ($S_{z} = - 1$) 
and (d) BM ($S_{z} = + 1$). 
}
\label{fig5} 
\end{figure}

Figure~\ref{fig5} shows the reflection and transmission coefficients when the quadrupolar spin mode is incident on the domain wall. 
Perfect reflection occurs in the low-momentum limit, 
but excitations with finite momentum pass through the domain wall. 
At higher energies, the Bogoliubov mode of $S_{z}=+1$ accounts for a substantial fraction of the transmitted flux. 
For perfect reflection in the low-momentum limit, 
we confirmed that the wave function vanishes in this limit. 
This is a conventional reflection property in the long-wavelength limit. 

The momentum transfer can be used to drive the ferromagnetic domain wall.
From the viewpoint of the momentum transfer that occurs when excitations are reflected, 
the spin wave in ferromagnets does not drive the ferromagnetic domain wall.
This is because this spin wave exhibits perfect transmission that is independent of the energy~\cite{Yan2011}. 
On the other hand, in the ferromagnetic spin-1 Bose--Einstein condensate, 
perfect reflection occurs for all three excitations: the transverse spin wave, the quadrupolar spin mode, 
and the Bogoliubov mode.
The first two excitations do not play a major role in driving the domain wall
because they exhibit perfect reflection in the low-momentum limit so that they involve very low momentum transfer. 
In contrast, the Bogoliubov mode exhibits perfect reflection outside of  the low-momentum limit (A and B in Fig.~\ref{fig4}). 
This relatively high momentum transfer of this mode is useful for driving the domain wall. 
Spin-transfer torque can also drive a ferromagnetic domain wall~\cite{Tatara2004,Yan2011}. 
However, the formalism used in this paper cannot directly demonstrate excitation-driven domain wall motion 
due to spin transfer as well as momentum transfer, 
because it involves only a small fluctuation about the condensate wave function. 
The subject of future studies is to directly demonstrate how excitation tunneling affects domain wall motion 
in the spin-1 Bose--Einstein condensate by including the nonlinear effect as studied in Ref.~\cite{Yan2011}.  

The simple domain wall shown in Figs.~\ref{fig1} and~\ref{fig2} can be mapped onto a domain wall soliton (phase separation) in binary mixtures of Bose--Einstein condensates~\cite{Coen2001} (see also (\ref{GPpm1})). 
In this regard, the phase separation can be experimentally controlled 
by changing the scattering length between same or different components through the Feshbach resonance~\cite{Papp2008,Tojo2010}. 
The term ``quadrupolar spin mode'' is specific to the spin-1 Bose--Einstein condensate; 
however, transmission and reflection properties equivalent to those shown in Figs.~\ref{fig4} and~\ref{fig5} 
are easier to obtain experimentally by employing an immiscible two-component Bose--Einstein condensate 
(e.g., $^{85}$Rb-$^{87}$Rb Bose--Einstein condensates~\cite{Papp2008} and Bose--Einstein condensates of $^{87}$Rb with two internal spin states~\cite{Tojo2010}). 

To create the localized Bogoliubov mode and localized spin-wave excitations, 
Bragg scattering and Raman scattering, whose light beams are shined over a razor edge, would be useful, respectively. 
After an excitation is scattered by the domain wall, if reflection occurs, excitations may propagate in the direction opposite to the incident excitation. Stern--Gerlach measurements can be used to determine their spin degrees of freedom. 
On the other hand, the movement and deformation of the domain wall after scattering can be observed by imaging the density
because the total particle density is low at the domain wall (Fig.~\ref{fig2}(b)). 

In this paper, to extract the essential physics, 
we considered a planar domain wall and one-dimensional scattering in a three-dimensional system. 
Nambu--Goldstone modes of a Bose--Einstein condensate are known to be reflected and refracted 
in the presence of a potential step~\cite{WatabeKato2008,WatabeKato2010}.  For junctions in Bose--Einstein condensates with equal densities, reflection and refraction are absent for both the Bogoliubov mode and a transverse spin wave. 
For junctions in the (anti-)ferromagnetic Heisenberg model, spin waves also transmit perfectly~\cite{Kato2012}. 
It will be interesting to extend the present one-dimensional scattering problem to a three-dimensional one
and to compare reflection and refraction properties with those reported in~\cite{WatabeKato2008,WatabeKato2010} and for (anti-)ferromagnets.  
The findings in this study will be helpful for determining the interface physics of superfluids and scattering processes through domain walls that have more complex structures than a planar wall. 
For scattering between low-lying modes and topological excitations, 
it is also a future problem to study Nambu-Goldstone mode scattering from a (half-)quantized vortex and a skyrmion in spinor Bose--Einstein condensates.

\section{Summary} 
We studied scattering between low-lying modes and a topological defect in a spin-1 Bose--Einstein condensate.
In particular, we report the structure of a ferromagnetic domain wall and the transmission properties of excitations through this domain wall.
We found that the domain wall and the tunneling properties of the transverse spin wave differ from those for a planar domain wall in the ferromagnetic Heisenberg model.
We also found that the transmission and reflection properties are strongly non-monotonic when the Bogoliubov mode is scattered from the domain wall.

\acknowledgements 
Authors thank D. Takahashi for useful discussions on zero-modes in Bose--Einstein condensates.  
SW thanks J. Ieda for discussions and pointing out references~\cite{Yan2011,Winter1961}. 
This work was supported by Grants-in-Aid for Scientific Research 
(Grant Nos. 20500044, 21540352, 22540412, 23104723, 23500056) from JSPS and MEXT, Japan. 

\appendix 
\section{How to determine the transmission and reflection coefficients}\label{Appendix}

Considering fluctuations in the condensate wave function ${\hat \Phi}(x,t)=\exp{(-i\mu t/\hbar)} ( {\hat \Phi(x)}+{\hat \phi}(x,t))$ 
and retaining terms to $O({\hat \phi}(x))$ in the Gross--Pitaevskii equation (\ref{eq.3}), 
we obtain equations for determining the tunneling properties of excitations. 
This appendix employs $\Phi_0 = 0$, which is consistent with the system we are studying in this paper. 
For a transverse spin wave (excitation with $S_{z}=0$ state), 
after applying the Bogoliubov transformation $\phi_{0} = u_{0} e^{-iEt/\hbar} - v_{0}^{*} e^{+iEt/\hbar}$, 
we obtain the following equation 
\begin{align}
E
\begin{pmatrix}
u_{0} \\ v_{0}
\end{pmatrix} 
= 
\begin{pmatrix}
h_{0} & -2c_{1} \Phi_{+1} \Phi_{-1} \\
2c_{1} \Phi_{+1}^{*} \Phi_{-1}^{*} & - h_{0} 
\end{pmatrix} 
\begin{pmatrix}
u_{0} \\ v_{0}
\end{pmatrix}, 
\label{eq5}
\end{align}
where $h_{0} = - \frac{\hbar^{2}}{2m}\partial_x^2 -\mu + c_{+} (|\Phi_{+ 1}|^{2} + |\Phi_{- 1}|^{2} )$. 
Amplitude transmission and reflection coefficients, $T_{0}$ and $R_{0}$, are obtained, 
by solving (\ref{eq5}) with the following boundary conditions: 
$
{\bf u}_{0}
= 
{\bf e}_{1}
e^{+ ikx} 
+ R_{0}
{\bf e}_{1}
e^{- ikx} 
+ A_{0} 
{\bf e}_{2}
e^{k x}
$
for $x = -\infty$
, and 
$
{\bf u}_{0}
= 
T_{0}
{\bf e}_{1}
e^{+ ikx} 
+ B_{0}
{\bf e}_{2}
e^{-k x}
$ 
for $x = +\infty$, 
where $\hbar k =  (2m E)^{1/2}$, 
${\bf u}_{0} \equiv (u_{0},v_{0})$, 
${\bf e}_{1} \equiv (1,0)$, and ${\bf e}_{2} \equiv (0,1)$. 
$A_{0}$ and $B_{0}$ are unknown coefficients of the damping solutions, 
which should be determined along with $T_{0}$ and $R_{0}$. 

On the other hand, as for the excitations with $S_{z} = \pm 1$ states, 
the usual Bogoliubov transformation $\phi_{\pm 1} = u_{\pm 1} e^{-iEt/\hbar} - v_{\pm 1}^{*} e^{+iEt/\hbar}$ leads to 
\begin{align}
E 
\begin{pmatrix}
u_{+1} \\ v_{+1} \\ u_{-1} \\ v_{-1}
\end{pmatrix}
= 
\begin{pmatrix}
h_{+} & -a_{+} & b_{+} & - c \\
a_{+}^{*} &  - h_{+} & c^{*} & - b_{+}^{*} \\
b_{-} & -c & h_{-} & -a_{-} \\ 
c^{*} & - b_{-}^{*} & a_{-}^{*} & - h_{-} 
\end{pmatrix}
\begin{pmatrix}
u_{+1} \\ v_{+1} \\ u_{-1} \\ v_{-1}
\end{pmatrix}, 
\label{4by4}
\end{align}
where $h_{\pm} = - \frac{\hbar^{2}}{2m}\partial_x^2 -\mu + 2 c_{+} |\Phi_{\pm 1}|^{2} + c_{-} |\Phi_{\mp 1}|^{2}$, 
$a_{\pm} = c_{+} \Phi_{\pm 1} \Phi_{\pm 1}$, $b_{\pm} = c_{-} \Phi_{\pm 1} \Phi_{\mp 1}^{*}$, and $c = c_{-} \Phi_{+1} \Phi_{-1} $. 
To determine the transmission and reflection coefficients, 
we impose the following boundary condition: 
\begin{align} 
\begin{pmatrix}
{\bf u}_{+1}\\ {\bf u}_{-1} 
\end{pmatrix}
= & 
\begin{pmatrix}
{\bf u}_{+1}^{\rm in} \\ {\bf u}_{-1}^{\rm in} 
\end{pmatrix}
+
R_{+1}
\begin{pmatrix}
{\bf e}_{1} \\ {\bf 0}
\end{pmatrix}
e^{- i k_{\rm Q} x}
+ A_{+1}
\begin{pmatrix}
{\bf e}_{2} \\ {\bf 0}
\end{pmatrix}
e^{ \kappa_{\rm Q} x}
\nonumber
\\
&
+R_{-1}
\begin{pmatrix}
{\bf 0} \\ \boldsymbol{\alpha}
\end{pmatrix}
e^{- i k_{\rm B} x}
+ A_{-1}
\begin{pmatrix}
{\bf 0} \\ \boldsymbol{\beta}
\end{pmatrix}
e^{ \kappa_{\rm B} x} (x = -\infty), 
\label{BC1}
\\
\begin{pmatrix}
{\bf u}_{+1}\\ {\bf u}_{-1} 
\end{pmatrix}
= & 
T_{+1}
\begin{pmatrix}
\boldsymbol{\alpha} \\ {\bf 0}
\end{pmatrix}
e^{+ i k_{\rm B} x}
+ B_{+1}
\begin{pmatrix}
\boldsymbol{\beta} \\ {\bf 0}
\end{pmatrix}
e^{ - \kappa_{\rm B} x}
\nonumber
\\
&
+
T_{-1}
\begin{pmatrix}
{\bf 0} \\ {\bf e}_{1}
\end{pmatrix}
e^{+ i k_{\rm Q} x}
+ B_{-1}
\begin{pmatrix}
{\bf 0} \\ {\bf e}_{2}
\end{pmatrix}
e^{ - \kappa_{\rm Q} x} (x = \infty), 
\label{BC2}
\end{align}
where ${\bf u}_{\pm 1}^{\rm T} \equiv (u_{\pm 1}, v_{\pm 1})$, 
$k_{\rm Q}$ and $\kappa_{\rm Q}$ are respectively the wavevectors of the quadrupolar spin mode and its damping solution, $k_{\rm B}$ and $\kappa_{\rm B}$ are respectively wavevectors of the Bogoliubov mode and its damping solution~\cite{kQ}. 
$T_{\pm 1}$ and $R_{\pm 1}$ are respectively the amplitude transmission and reflection coefficients for propagating modes, 
and $A_{\pm 1}$ and $B_{\pm 1}$ are coefficients for damping modes, which should be determined in this problem.
For the Bogoliubov mode incident to the domain wall, we set 
$({\bf u}_{+1}^{\rm in}, {\bf u}_{-1}^{\rm in}) 
= ({\bf 0}, \boldsymbol{\alpha}) \exp{(i k_{\rm B} x)}$. 
On the other hand, for the incident quadrupolar spin mode, 
we set 
$({\bf u}_{+1}^{\rm in}, {\bf u}_{-1}^{\rm in}) 
= ({\bf e}_{1} ,  {\bf 0}) \exp{(i k_{\rm Q} x)}$.

\section{Proof of perfect reflection of transverse spin wave}\label{Appendix2}

In this appendix, we prove and discuss perfect reflection of the transverse spin wave mode in the low-energy limit.
We will use the dimensionless form, where the wave function, energy, and length are scaled by 
$\sqrt{\rho}$, $c_+ \rho$ and $\xi = \hbar/\sqrt{mc_+  \rho}$. 
Two functions $S_{\pm} \equiv u_0 \pm v_0$ obey the following equation 
\begin{align}
h_{0}^{(\pm )} S_{\mp} \mp i 2c_1 {\rm Im}(\Phi_{+1}\Phi_{-1}) S_{\pm} = E S_{\pm}
\end{align}
where $h_{0}^{(\pm)} \equiv h_0 \pm 2 c_{1} {\rm Re}(\Phi_{+1}\Phi_{-1})$. 
Since $\Phi_{\pm 1}$ are real in this study, 
we have 
\begin{align}
h_{0}^{(\pm )} S_{\mp} = E S_{\pm}. 
\label{Spm}
\end{align}
The flux density, which is independent of  $x$, is given by 
$J \propto {\rm Im}[S_{+}^{*}\partial_x S_{+} + S_{-}^{*} \partial_x S_{-}] $. 
For the tunneling problem, the boundary condition for $x = -\infty$ is given by 
$S_{\pm} = e^{ikx} + R_0 e^{-ikx} \pm A_0 e^{kx}$, 
and that for $x = +\infty$ is given by 
$S_{\pm} = T_0 e^{ikx} \pm B_0 e^{-kx}$. 
The transmission and reflection coefficients $\tau$ and $r$ are now determined as $\tau = |T_0|^2$ and $r = |R_0 |^2$, respectively. 

To study the low-energy properties, we use the following expansion 
\begin{align}
(T_0,R_0,A_0,B_0) = \sum\limits_{n=0} k^{n} (T_{0}^{(n)}, R_{0}^{(n)}, A_{0}^{(n)}, B_{0}^{(n)}). 
\end{align}
For $k |x| \ll 1$, the boundary condition for $x \ll - 1$ is given by 
$S_{\pm} = 1 + R_{0}^{(0)} \pm A_{0}^{(0)} + k (R_{0}^{(1)} \pm A_{0}^{(1)}) 
+ kx (i - i R_{0}^{(0)} \pm A_{0}^{(0)}) + \cdots $, 
and that for $x \gg 1$ is given by 
$S_{\pm} = T_{0}^{(0)} \pm B_{0}^{(0)} + k (T_{0}^{(1)} \pm B_{0}^{(1)}) + kx (i T_{0}^{(0)} \mp B_{0}^{(0)}) + \cdots $. 

We also determine $S_{\pm}$ from the equation (\ref{Spm}). 
We expand the functions $S_{\pm}$ as 
$S_{\pm} = \sum_{n=0} k^{n} S_{\pm}^{(n)}$. 
Then, $h_0^{(\mp)} S_{\pm}^{(0,1)} = 0$ follows. 
Since one of the zero mode solutions for (\ref{eq5}) is given by $(u_0, v_0) = (\Phi_{+1}, \Phi_{-1}^*)$, 
$S_{\pm, \rm I}^{(0,1)} = \Phi_{+1} \pm \Phi_{-1} \equiv \phi_\pm $ follow, 
where we took $\Phi_{\pm 1}$ to be real. 
The other solutions are obtained as 
$S_{\pm, \rm II}^{(0,1)} (x) = \phi_{\pm} (x) \int_{0}^{x} dx' \phi_{\pm}^{-2} (x') $. 
The behaviors for $|x| \gg 1$ are given by 
$(S_{+, \rm I}^{(0,1)}, S_{-, \rm I}^{(0,1)}) = (1,  {\rm sgn}(x))$ and 
$(S_{+, \rm II}^{(0,1)}, S_{-, \rm II}^{(0,1)}) = ({\rm sgn}(x) \gamma_{+} + x,  \gamma_{-} + |x| )$, 
where $\gamma_{\pm} \equiv \int_{0}^{\infty} dx (\phi_{\pm}^{-2} - 1)$. 
Here, we used $\phi_{\pm} (- x) = \pm \phi_{\pm} (x)$. 
We examine solutions linear in $k$, which are given by 
$S_{\pm} (x) 
= 
C_{\pm, \rm I}^{(0)} S_{\pm, \rm I}^{(0)} + C_{\pm, \rm II}^{(0)} S_{\pm, \rm II}^{(0)} 
+ k (C_{\pm, \rm I}^{(1)} S_{\pm, \rm I}^{(1)} + C_{\pm, \rm II}^{(1)} S_{\pm, \rm II}^{(1)} )
$. 
By comparing the coefficients of $k$ and $x$, 
we obtain 
\begin{align}
(R_0^{(0)}, T_0^{(0)}, A_0^{(0)}, B_0^{(0)}) = & (- i , 0, 0, 1 - i), \\ 
\pm C_{\pm ,\rm I}^{(0)} = \mp C_{\pm,\rm II}^{(1)} = & 1- i, \\
C_{\pm,\rm II}^{(0)}  = & 0. 
\end{align}
As a result, the excitation shows perfect reflection in the low-energy limit (i.e., $|R_0^{(0)}|^2 = 1$). 
We also note that the wave functions show $\lim\limits_{E\rightarrow 0} S_{\pm} =  \pm (1-i) \phi_{\pm}$, 
that is 
\begin{align}
\lim\limits_{E\rightarrow 0} (u_0, v_0)/ (1-i)  = ( \Phi_{-1}, \Phi_{+1} ). 
\end{align} 
The spin density fluctuation $\delta {\bf F}$, defined here as the spin density ${\bf F} = \hat \Phi^\dag {\bf S} \hat \Phi$ 
within the first order of $\phi_0$, is given by $\delta F_{x,y} \propto  F_{z}$, where $F_z = \Phi_{+1}^2 - \Phi_{-1}^2$. 

As for the usual perfect reflection in quantum mechanics, the phase of the amplitude reflection coefficient $R_0$ in the low-energy limit leads to $\pi$ and the amplitude of the wave function vanishes everywhere. 
However, in this case, the phase of the amplitude reflection coefficient $R_0$ in the low-energy limit is $3\pi/2$
and the excitation wave function remains in the low-energy limit when perfect reflection occurs. 
Perfect reflection and a non-vanishing wave function coexist in the low-energy limit for the following reason. 
The transverse spin mode in this system has a damping solution whose damping length is $1/k$. 
In the low-energy limit, this damping length becomes infinite and the damping mode can smoothly connect with the condensate wave function, which is uniform for $|x| \gg 1$. 
Consequently, the propagating wave mode can vanish in this limit.

The coexistence of perfect reflection and a non-vanishing wave function is also observed in the scalar Bose--Einstein condensate. 
In particular, we find perfect reflection in the low-energy limit 
when the Bogoliubov excitation is incident on the region 
where the condensate density is absent due to the presence of a potential step that exceeds the chemical potential~\cite{Tsuchiya2009,TakahashiarXiv2009}. 
If we regard the excitation in the low-energy limit as the phase mode of the condensate, 
perfect reflection in the low-energy limit is natural 
because a region with no condensate density cannot support the phase mode. 
On the other hand, in the present case, the spin wave mode in the
low-energy limit, which is related to the rotation symmetry
of the spin space, can propagate on both sides of the domain wall, 
so that perfect reflection of this mode originates from different causes. 
A zero-value of the order parameter at the domain wall is specific to the present system, 
and is suggestive of perfect reflection in the low-energy limit. 

However, perfect reflection of this spin mode does not originate from the zero-value of the spin density at the domain wall. 
To demonstrate this, 
we assume that $\lim\limits_{x\rightarrow \pm \infty} (\Phi_{+1}, \Phi_{-1}) = (1,0)$;
however, the system has a region where $(\Phi_{+1}, \Phi_{-1}) = (0,1)$ around $x = 0$. 
For the tunneling problem, the boundary conditions for excitations are unaltered. 
Solutions $S_{\pm, \rm I}^{(0,1)}$ and $S_{\pm, \rm II}^{(0,1)}$ are formally given in the same form as those derived above. 
However, they have different behaviors, which are given by 
$S_{\pm, \rm I}^{(0,1)} = 1$ and $S_{\pm, \rm II}^{(0,1)} = {\rm sgn}(x) \gamma_{\pm} + x$, 
where $\gamma_{\pm} \equiv \int_{0}^{\infty} dx (\phi_{\pm}^{-2} - 1)$. 
Following the same procedures shown above, 
we obtain 
\begin{align}
(R_0^{(0)}, T_0^{(0)}, A_0^{(0)}, B_0^{(0)}) = & (0 , 1, 0, 0), \\
( C_{\pm, \rm I}^{(0)}, C_{\pm, \rm II}^{(0)}, C_{\pm, \rm II}^{(1)}) = & (1,0,i). 
\end{align}
As a result, we have perfect transmission in the low-energy limit (i.e., $|T_0^{(0)}|^2 = 1$), 
and $\lim_{E\rightarrow 0} (u_0, v_0) = (\Phi_{+1}, \Phi_{-1})$. 
The spin density fluctuation is given by $(\delta F_x, \delta F_y) \propto (F_z, 0)$. 
In the present case, we can find points where $\Phi_{+1} = \Phi_{-1}$, which leads to $F_z = 0$. 
However, the absence of the amplitude of the order parameter at the domain walls 
is irrelevant to the tunneling properties in the low-energy limit. 
In a scalar Bose--Einstein condensate, the Bogoliubov excitation can pass through the dark soliton~\cite{Kovrizhin2001}. 
This is another example in which perfect transmission occurs in a system where the amplitude of the order parameter is absent.

\section{Details of Zero-Mode Tunneling through a Soliton}\label{Appendix3}

In this appendix, we discuss effect of zero-energy modes on the tunneling properties 
by focusing on the phase and density fluctuations. 

First, by comparing (\ref{4by4}) for $E=0$ with (\ref{GPpm1}) and its derivative with respect to $x$, 
we find two simple zero-energy mode solutions for (\ref{4by4}), which are given by 
\begin{align}
(u_{+1}, v_{+1}, u_{-1}, v_{-1}) = & (\Phi_{+1},\Phi_{+1}^{*},\Phi_{-1},\Phi_{-1}^{*}),
\label{ZeroModePhase}
\\
(u_{+1}, v_{+1}, u_{-1}, v_{-1}) = & \partial_{x} (\Phi_{+1},- \Phi_{+1}^{*}, \Phi_{-1}, - \Phi_{-1}^{*}), 
\label{ZeroModeX}
\end{align} 
where we omit the normalization factors. 
If we add an external potential term to (\ref{GPpm1}) and (\ref{4by4}) by replacing $\mu$ with $\mu - V(x)$, 
the first solution (\ref{ZeroModePhase}) holds, whereas the second solution (\ref{ZeroModeX}) does not. 
(\ref{ZeroModeX}) yields a zero mode solution to (\ref{4by4}) only when the system is uniform. 
This is easy to understand when we consider (spontaneously) broken symmetries. 
As for the broken gauge symmetry, when the condensate wave function is given by $\Phi (x) = A(x)\exp{[i\theta (x)]}$, 
the function $A(x)\exp{\{i[\theta(x) + d\theta]\}} \simeq \Phi (x) + i\Phi(x) d\theta$ is another solution of (\ref{GPpm1}). 
The factor proportional to $d\theta$ leads to (\ref{ZeroModePhase}). 
(Here, we omit the subscript $\pm 1$; the discussion in the present paragraph also holds for a scalar Bose--Einstein condensate.)
Thus, the first solution (\ref{ZeroModePhase}) is the zero mode related to the broken gauge symmetry and it does not contribute the density fluctuation.
(When the condensate wave function $\Phi$ fluctuates as $\Phi + \phi$, 
the density fluctuation is given by $| \Phi + \phi |^2 - |\Phi|^2 = 2 {\rm Re}[ \Phi^* \phi] + {\mathcal O}(\phi^2)$. 
In the low-energy limit, $\phi = u - v^*$ and the solution (\ref{ZeroModePhase}) is not related to the density mode.) 
On the other hand, in the absence of an external potential, 
$\Phi(x + dx) \simeq \Phi(x) + \partial_{x} \Phi (x) dx$ is another solution of (\ref{GPpm1}). The term proportional to $dx$ leads to (\ref{ZeroModeX})~\cite{ZeroHeisenberg}. 
Thus, the second solution (\ref{ZeroModeX}) is the zero mode related to the broken translational symmetry 
due to the presence of the soliton. 
This gives a finite density fluctuation and the matrix element of the density spectral function is given by $F(x) \propto (\partial_{x} \rho(x) )^{2}$ where $\rho(x) = |\Phi(x)|^{2}$. 
The profile in Fig.~\ref{fig4}(e) pertains to the present discussion. 
To summarize, when the condensate exhibits solitary behavior, 
the density fluctuation does appear in the soliton. 

\begin{figure}[tbp]
\includegraphics[width=9cm]{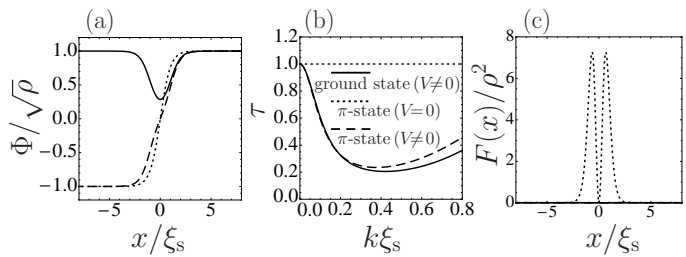}
\caption{
Transmission and reflection properties of the Bogoliubov excitation in a scalar Bose--Einstein condensate. 
We consider three cases: (i) the ground state in the presence of the barrier $(V\neq 0)$ (solid lines in each panel), 
(ii) $\pi$-state in the absence of the barrier $(V = 0)$ (dotted lines in each panel), 
and (iii) $\pi$-state in the presence of the barrier $(V \neq 0)$ (dashed lines in each panel). 
(a) the condensate wave function as a function of $x$, (b) momentum-dependent transmission probability $\tau = |T|^{2}$, 
and (c) the matrix element of the density spectral function in the low-energy regime $k\xi_{\rm s}= 10^{-2}$. 
$\xi_{\rm s}$ is the healing length, which is given by $\xi_{\rm s} = \hbar /\sqrt{mg\rho}$. 
For cases (i) and (iii), we used $V(x) = 2g\rho \exp(-x^{2}/\xi_{\rm s}^{2})$. 
}
\label{fig6} 
\end{figure} 
We here discuss the absence of reflection of the Bogoliubov mode in the low-energy limit from the domain wall shown in Fig.~\ref{fig4}. Starting with a scalar Bose--Einstein condensate is instructive for understanding the characteristic properties of excitation tunneling through the domain wall. 
In the usual manner and using the usual notation, the condensate wave function is obtained by solving the Gross--Pitaevskii equation 
\begin{align}
\left [ - \frac{\hbar^{2}}{2m}\partial_x^2 + V(x) - \mu + g |\Phi (x)|^{2} \right ]\Phi (x) = 0. 
\label{GPscalar}
\end{align}
The Bogoliubov excitation is described by 
\begin{align}
E
\begin{pmatrix}
u \\ v 
\end{pmatrix} 
= 
\begin{pmatrix}
h & - g \Phi^{2} \\
g (\Phi^{*})^{2} & - h
\end{pmatrix} 
\begin{pmatrix}
u \\ v
\end{pmatrix}, 
\label{BogoScalar}
\end{align}
where $h = - \hbar^{2} / (2m) \partial_x^2 + V(x) - \mu + 2g |\Phi (x)|^{2}$. 
The chemical potential $\mu$ and 
the boundary condition ${\bf u} \equiv (u, v)$ for the tunneling problem are respectively given by 
$\mu = g\rho \equiv g|\Phi(x\rightarrow 0)|^{2}$ and  
\begin{align} 
{\bf u} 
= & 
\boldsymbol{\alpha} e^{+ i k_{\rm B} x}
+R \boldsymbol{\alpha}
e^{- i k_{\rm B} x}
+ A \boldsymbol{\beta} 
e^{ \kappa_{\rm B} x} (x = -\infty), 
\label{BC1}
\\
{\bf u}
= & 
T \boldsymbol{\alpha} e^{+ i k_{\rm B} x}
+ B \boldsymbol{\beta}
e^{ - \kappa_{\rm B} x} 
(x = \infty), 
\label{BC2}
\end{align}
where $k_{\rm B}$, $\kappa_{\rm B}$, $\boldsymbol{\alpha}$ and $\boldsymbol{\beta}$ are given in \cite{kQ} (where $c_{+}$ is replaced by $g$). 
$T$ and $R$ are respectively amplitude transmission and reflection coefficients for propagating modes
and $A$ and $B$ are the respective coefficients for damping modes, which should be determined in this problem. 
We consider three cases below. The first case is the ground state in the presence of the barrier $V\neq 0$,
where the boundary condition of the condensate wave function is given by $\Phi(x\rightarrow \pm \infty) = + \sqrt{\rho}$. 
This case was studied as an anomalous tunneling problem~\cite{Kagan2003}. 
The results are shown by the solid lines in each panel of Fig.~\ref{fig6}. 
In this case, the Bogoliubov mode perfectly tunnels through the barrier in the low-energy limit (Fig.~\ref{fig6}(b)) and the density mode is suppressed (Fig.~\ref{fig6}(c)). 
The second case is the $\pi$-state in the absence of a barrier ($V= 0$) 
where the boundary condition of the condensate wave function is given by $\Phi(x\rightarrow \pm \infty) = \pm \sqrt{\rho}$. 
This case was considered in~\cite{Kovrizhin2001}. 
The results are shown by the dotted lines in each panel of Fig.~\ref{fig6}. 
In this case, the Bogoliubov mode perfectly tunnels through the soliton, independent of the energy (Fig.~\ref{fig6}(b)).
In contrast to the previous case, density fluctuation appears (Fig.~\ref{fig6}(c)). 
We here examine whether the Bogoliubov mode transmits through the barrier in the low-energy limit when the system is in the $\pi$-state. 
The results are shown by the dashed lines in each panel of Fig.~\ref{fig6}. 
In the low-energy limit, the Bogoliubov excitation in the $\pi$-state perfectly tunnels through the barrier at the center of the soliton. 
In this case, the potential barrier suppresses the zero mode of the density. 
From these three cases, we find that the density fluctuation is irrelevant to perfect tunneling 
through a soliton in a scalar Bose--Einstein condensate. 
Since the presence of the zero mode of the phase is related to the phase coherence of the condensate, 
this coherence is important for perfect tunneling of the low-energy Bogoliubov mode in a scalar Bose--Einstein condensate. 

\begin{figure}[tbp]
\includegraphics[width=7.3cm]{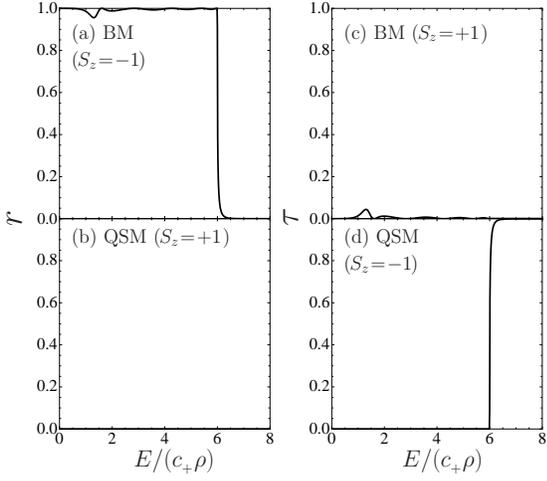}
\caption{
Transmission and reflection properties of excitations through the domain wall 
deformed by the potential barrier $V(x)$ when the Bogoliubov mode is incident. 
Reflection probabilities of (a) Bogoliubov mode (BM) ($S_{z} = - 1$) 
and (b) quadrupolar spin mode (QSM) ($S_{z} = + 1$). 
Transmission probabilities of (c) BM ($S_{z} = + 1$) 
and (d) QSM ($S_{z} = - 1$). 
The coupling constant is chosen as $c_{1} = -3c_{+}$ and we used $V(x) = 2c_{+} \rho \exp(-x^{2}/\xi^{2})$. 
We here took the center of the domain wall (i.e., $x=0$) as the center of the potential barrier. 
}
\label{fig7} 
\end{figure}

We now examine the case for the domain wall of a spin-1 Bose--Einstein condensate. 
We add an external potential to the center of the domain wall, which suppresses the zero mode of the density, 
to investigate whether the perfect conversion of energy flux from $S_{z} = -1$ Bogoliubov mode to $S_{z} = +1$ Bogoliubov mode occurs only through the zero mode of the phase. 
We add an external potential term to (\ref{GPpm1}) and (\ref{4by4}) by replacing $\mu$ with $\mu - V(x)$. 
The results are shown in Fig.~\ref{fig7}. 
We find that perfect reflection occurs in the low-energy limit when the Bogoliubov mode is injected into the domain wall deformed by the barrier. 
This result contrasts with that in the $\pi$-state of a scalar Bose--Einstein condensate. 
We confirmed that in the presence of the barrier, the matrix element of the density spectral function near the domain vanishes as the energy decreases.
This result suggests that the localized density mode plays an important role in converting the energy flux of a low-energy Bogoliubov mode propagating in one Bose--Einstein condensate into that in the other condensate.

We also confirmed that reflection is absent in the low-energy limit when the barrier is located in a region where the condensate is uniform (i.e., far from the interface between condensates). 
In this case, the potential barrier breaks the translational invariance so that the zero mode solution is not given by $\partial_{x} \Phi_{\pm1}$. 
However, we find the same behavior of the matrix element of the density spectral function as shown in Fig.~\ref{fig4}(e). 
The breaking of the translational invariance by a potential barrier far from a domain wall 
is irrelevant to the appearance of the localized density mode near the domain wall and to the absence of reflection.


\begin{thebibliography}{99}
\bibitem{QZ} 
For transmission and reflection properties of excitations, 
the quadratic Zeeman effect affects only the transverse spin mode, but this does not affect the qualitative result.
\bibitem{kQ} 
Parameters are given by 
$\hbar k_{\rm Q} \equiv [2m (E + 2c_{1} \rho)]^{1/2}$, 
$\hbar \kappa_{\rm Q} \equiv [2m (E - 2c_{1} \rho)]^{1/2}$, 
$\hbar k_{\rm B} \equiv [ 2m (WE - c_{+}\rho) ]^{1/2}$, 
$\hbar \kappa_{\rm B} \equiv [ 2m (WE + c_{+}\rho) ]^{1/2}$, 
$\alpha \equiv [(W+1)/2 ]^{1/2}$, and $\beta \equiv [(W-1)/2 ]^{1/2}$, 
where $W \equiv \left [  1 + (c_{+}\rho/E)^{2} \right ]^{1/2}$. 
We also define $\boldsymbol{\alpha} \equiv (\alpha, \beta)$ and 
$\boldsymbol{\beta} \equiv (\beta, -\alpha)$.
\bibitem{ZeroHeisenberg}
This type of zero mode is also found in the ferromagnetic Heisenberg model. 
As for the domain wall solution to (\ref{Hspin}) given by $(\sigma_{x}, \sigma_{y}, \sigma_{z}) = (0,\sigma/\cosh(x/\lambda),\sigma\tanh(x/\lambda)) 
\equiv \sigma (0, \cos\theta_{0}(x), \sin\theta_{0}(x))$, 
the zero mode solution $\varphi_{0} (x) = 1/\cosh(x/\lambda)$ is related to $\theta_{0}(x)$ through $\varphi_{0} (x) = - \lambda \partial_{x} \theta_{0}(x)$~\cite{TataraReview}.
\end{thebibliography}
\end{document}